%% file: kaonatkloe_4_arxiv.tex
\documentclass{appolb}
\usepackage{graphicx}
\usepackage{enumerate}
\include{definition}

\begin{document}
\title{
Kaon physics with the KLOE detector
\thanks{To appear on a special issue of Acta Physica Polonica B}%
}
\author{C.~Bloise, E.~De~Lucia, A.~De~Santis, P.~De~Simone
\address{Laboratori Nazionali di Frascati dell'INFN, Frascati, Italy}\\
\vskip 0.5cm
E.~Czerwi\'nski, A.~Gajos, D.~Kami\'nska, P.~Moskal, M.~Silarski
\address{Institute of Physics, Jagiellonian University, Cracow, Poland}
\vskip 0.5cm
A. Di Domenico 
\address{Dipartimento di Fisica, Sapienza Universit\`a di Roma, Roma, Italy\\
INFN sezione di Roma, Roma, Italy}
\vskip 0.5cm
A. Passeri
\address{INFN sezione di Roma Tre, Roma, Italy}
\vskip 0.5cm
W.~Wi\'slicki
\address{National Centre for Nuclear Research, Warsaw, Poland}
\\
\address{on behalf of the KLOE-2 collaboration}
}
\maketitle
\begin{abstract}
In this paper we discuss the recent
finalized analyses
by 
the KLOE experiment at DA$\Phi$NE:
the CPT and Lorentz invariance test with entangled
$\kzero \kzerob$ pairs, and the precision measurement of
the 
branching fraction of the decay \Kp3pig . 
\\We also present the status of an ongoing analysis aiming to precisely measure
the $\kpm$ mass.
\end{abstract}


\section{Introduction}
The KLOE experiment at DA$\Phi$NE, the Frascati $\phi$--factory, 
measured all the most relevant branching ratios of 
$\kl$, $\ks$, and $\kpm$ mesons, their
lifetimes, the $\kk \rightarrow \pi$ form factors, and several parameters and observables
allowing to perform
CP, CPT symmetries tests, and
a verification of 
the validity of Cabibbo unitarity and lepton universality (a review can be found in Ref.\cite{nuovocimento}).
\par
In this paper we focus the discussion on the two most recent results on kaon physics:
a CPT and Lorentz invariance test with entangled
$\kzero \kzerob$ pairs, and the precision measurement of
the 
branching fraction of the decay \Kp3pig . 
Among the ongoing analyses there are:
\begin{enumerate}[(a)]
\item
the measurement of the charged kaon mass, 
\item
the study of semileptonic $\ks$ decays and CPT symmetry tests, 
\item
the study of rare $\ks$ decays with an improved $\ks$ tagging technique; 
\item
the direct T--symmetry test with entangled neutral kaons; 
\end{enumerate}
The status of analysis (a) is also presented in
this paper, while (b), (c) and (d)
are the subjects of separate contributions in this issue \cite{daria,michal,alek}.

\section{CPT and Lorentz--invariance test}
\label{sec-2}
 In year 2014 KLOE obtained the best sensitivity ever
reached in the quark sector on testing CPT and Lorentz invariance, based on
1.7 fb$^{-1}$ of integrated luminosity~\cite{Babusci:2013gda}. 
The test was performed on the entangled neutral kaon pairs, in the
$\phi\to\ks\kl\to\pi^+\pi^- \pi^+\pi^- $ final state, studying
the interference pattern as a function of sidereal time and particle direction in
celestial coordinates. 

\noindent
Data reduction is based on two decay vertices with only two tracks
each. For each vertex several requirements, 
invariant mass $|m_\mathrm{rec}-m_\mathrm{K}|<5$~MeV, missing mass  
 $\sqrt{E_\mathrm{miss}^2+|\vec{p}_\mathrm{miss}|^2}<10$~MeV,
 -50~MeV$^2<m^2_\mathrm{miss}<10$~MeV$^2$, and kaon momenta
 compatible with the 2--body decay hypothesis, have been applied for
 the selection of a pure, well reconstructed data sample.
 The missing momentum is obtained from the analysis of Bhabha
 scattering events in the same run. 
A global likelihood function is built in order to kinematically
constrain the event and to improve on the event reconstruction quality
and in particular on the vertex resolution.

\noindent
Main background source is the kaon regeneration on the spherical beam
pipe, suppressed by selecting events with both \ks\ and \kl\ decaying
inside the beam pipe, giving at the end of the analysis chain a contamination at 2--3\% level.
Background contributions and analysis selection efficiencies are derived from MC simulation 
to which corrections from real data are applied. 
Full analysis chain was repeated several times varying all the cuts
for the evaluation of the systematic uncertainties. The sum in quadrature of all the 
effects ranges  between 30\% and 40\% 
of the statistical error.

\begin{figure}[ht]
\centering
\vskip -0.3cm
\includegraphics[width=.9\textwidth]{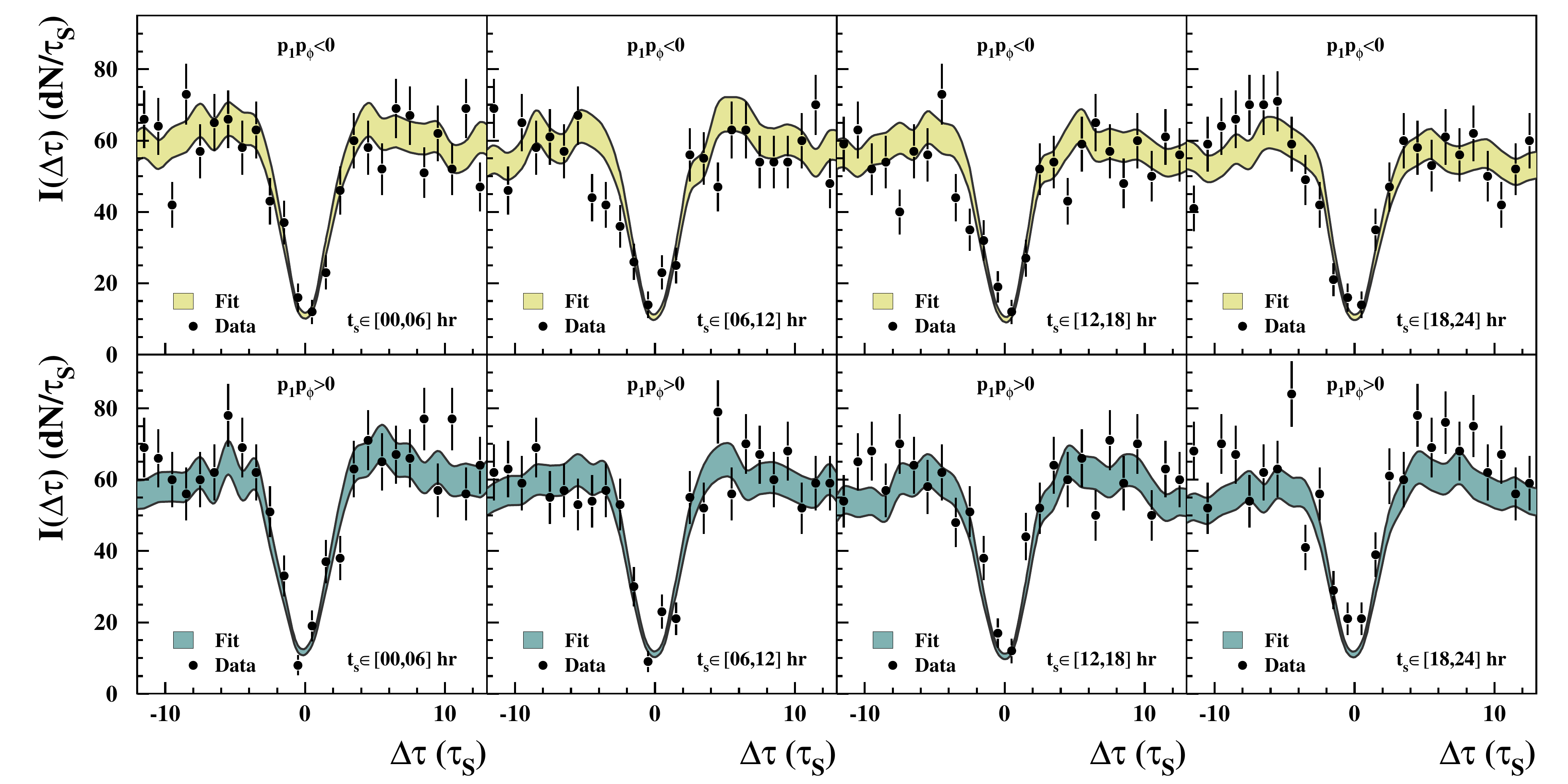} 
\vskip -0.3 cm
\caption{Fit to the distribution of the difference in proper time of neutral
  kaons as a function of sidereal time and particle direction in
  celestial coordinates~\cite{Babusci:2013gda}.}
\label{fig:plbplot}
\end{figure}

\noindent
The CPT test is based on 
the distribution of the difference $\Delta \tau$ in the decay proper time between
neutral kaons. Due to the fully--destructive quantum interference
at $\Delta\tau$=0, the  distribution is very sensitive to
CPT--violating effects, especially for the selected decays inside 
the beam pipe. Effects of CPT and Lorentz--invariance breaking in the low
energy regime relating with
modifications of the space--time structure at the Planck scale, are
described by the Standard Model Extension,
SME~\cite{Kostelecky:2008ts}, providing the parametrization: 
 \begin{center}
\vskip -0.8cm
    \begin{equation}
       \delta_k \sim i \sin \phi_\mathrm{SW} e^{i \phi_\mathrm{SW}} \gamma_\mathrm{K} 
                                              (\Delta
                                              a_0-\vec{\beta_\mathrm{K}}\cdot
                                              \Delta{\vec{a}})/\Delta
                                              m  \nonumber
     \end{equation} 
\end{center}
where $\gamma_\mathrm{K}$ and $\vec{\beta_\mathrm{K}}$ are the kaon $\gamma$--factor 
and velocity in the laboratory frame, 
$\phi_\mathrm{SW}=\arctan (2\Delta m/\Delta\Gamma)$ is the 
{\it superweak} phase, $\Delta m = m_\mathrm{L}-m_\mathrm{S}$, 
$\Delta \Gamma = \Gamma_\mathrm{S}-\Gamma_\mathrm{L}$ are the mass and width 
differences for the neutral 
kaon mass eigenstates and $\Delta a_{\mu}$ are the four CPT--violating coefficients. 
In this context, the intensity of kaon decays as a function of $\Delta\tau$ is expressed
by:
\begin{equation}
\label{eq:intro:timeevo} 
  I(\Delta\tau)   =  
 C_{12} ~ e^{-\Gamma |\Delta\tau|} \Big[ 
    |\eta_{1}|^2 e^{\frac{\Delta\Gamma}{2} \Delta\tau}  + 
    \,\,|\eta_{2}|^2 e^{-\frac{\Delta \Gamma}{2}\Delta\tau}
    -2 \Re\Big(\eta_{1}\eta_{2}^*e^{-i\Delta m \Delta\tau} \Big) \Big] 
\end{equation}
with $\Gamma = (\Gamma_\mathrm{S}+\Gamma_\mathrm{L})/2$,
\begin{eqnarray}
\eta_{1,2} &\simeq& \epsilon_k - \delta_k(\overrightarrow{p}_{1,2},t_{1,2}) ~,\nonumber \\
C_{12}&\simeq&\frac{ 1 }{2 (\Gamma_\mathrm{S}+\Gamma_\mathrm{L})} 
|\braket{f_1}{T|\ks}\braket{f_2}{T|\ks}|^2~.\nonumber
\end{eqnarray}  
It is used to fit the experimental distributions shown in fig.~\ref{fig:plbplot}.
The results~\cite{Babusci:2013gda} are the most sensitive measurements in the quark sector of SME:
\begin{center}
{$\Delta a_o = (-6.0 \pm 7.7_{stat} \pm 3.1_{syst})\times 10^{-18} $ GeV \\
$\Delta a_x = (\,\,\,0.9 \pm 1.5_{stat} \pm 0.6_{syst})\times 10^{-18}$ GeV \\
$\Delta a_y = (-2.0 \pm 1.5_{stat} \pm 0.5_{syst})\times 10^{-18} $ GeV \\
$\Delta a_z = (\,\,\,3.1 \pm 1.7_{stat} \pm 0.5_{syst})\times 10^{-18}$ GeV }
\end{center}

\noindent
For comparison, the accuracy reached by similar
measurements in B and D systems is of  $O(10^{-13})$~GeV~\cite{Kostelecky:2008ts}.
\section{The \Kp3pig ~ branching fraction }
\label{BRm}
The measurement of the branching fraction (BR) of \Kp3pig 
 is based on a sample of 
$\sim$17 million tagged $K^+$  mesons.
The $\kp$ are tagged by the reconstruction of 2--body $K^-$
decays that provide the normalization sample for the   
measurement of the absolute BR. 
\noindent
The analysis procedure consists of 
i) the selection of K$^+$ candidates 
(tagging procedure) by the identification 
of K$^- \to \pi^- \pi^0$ and K$^- \to \mu^- \nu$ samples, independently 
treated; ii) the reconstruction of the K$^+$ path from the kinematical 
constraints given by the K$^-$ momentum and $\phi$ momentum 
(from Bhabha--scattering events);
iii) the backward extrapolation  
of any charged track not belonging to the K$^-$ decay chain; iv) the 
reconstruction of the K$^+$ decay vertex radial position ($\rho_{xy}$) 
and the closest--approach distance (CAd$_i$) between the track and the
K$^+$ flight path; 
v) the selection of events with at least two tracks with CAd$_i \le$~3~cm 
and $\rho_{xy} \le$~26~cm,  outside the drift chamber (DC) sensitive
volume (for a better control 
of systematics from tagging procedure); 
vi) the measurement of the missing--mass distribution, 
M$^{2}_{miss}$ = ($\Delta$E$_{K^+ -\pi\pi}$)$^2$ - $|${\bf
  $\Delta$P}$_{K^+ -\pi\pi}|^2$. 
 
\noindent
The analysis is fully inclusive of radiative decays.

\noindent
The dependence of the K$^+ \to \pi^+ \pi^+ \pi^-$ signal sample on the tag selection (tag--bias) has been derived from Monte Carlo
simulations and used as a correction factor in the BR evaluation.  
The relating systematic error is estimated by changing the selection
criteria of the 2--body kaon decay
reconstruction.
\noindent
The other sources of systematic errors are 
the analysis cuts and the uncertainty on the kaon
lifetime \cite{Ambrosino:2007xz}. Results of their evaluation are
shown in table \ref{tabsys}.
\noindent
\begin{table}{
\caption{Summary table of the fractional systematic uncertainties~\cite{Babusci:2014hxa}.\label{tabsys}}
\begin{tabular}{ccc}
\hline\hline
\bf {Source of systematic uncertainties} &  $K^-_{\mu 2}$ tags $~(\%$)  & $K^-_{\pi 2}$ tags $~(\%$)  \\
\hline
DCA, DCA$_{12}$, cos$(\theta_{12})$ cuts & 0.52  & 0.41 \\
p$^*_{m_{\pi}}$ cut  & 0.08  & 0.11 \\
m$^2_{miss}$ cut &  0.05 & 0.14 \\
fiducial volume  &  0.11 & 0.10 \\
selection efficiency estimate  &  0.16 & 0.16  \\
tag bias & 0.16 & 0.32 \\
$K^{\pm}$ lifetime & 0.12 & 0.12 \\ 
\hline
Total fractional systematic uncertainty  & 0.60 & 0.59  \\
\hline\hline
\end{tabular}}
\end{table}
In a sample of 12,065,087 (5,171,239) $K^- \raE \mu^- \bar{\nu}
$ ($K^- \raE \pi^- \pi^0 $ ) tagging
events we found 
$N_{K \raE 3\pi} = 48,032 \pm 286$ ($20,063 \pm 186$) signal events,
corresponding to the fully consistent measurements of the absolute branching fractions:
\vskip -0.5cm
\begin{equation}
BR(K^+ \raE \pi^+\pi^-\pi^+(\gamma))|_{Tag K_{\mu2}} = 0.05552 \pm
0.00034_{stat} \pm 0.00034_{syst} \, 
\end{equation}
\vskip -0.8cm
\begin{equation}
BR(K^+ \raE \pi^+\pi^-\pi^+(\gamma))|_{Tag K_{\pi2}}  = 0.05587 \pm
0.00053_{stat} \pm 0.00033_{syst}. 
\end{equation}
\vskip -0.2cm

\noindent
The average, $BR(K^+ \raE \pi^+\pi^-\pi^+(\gamma)) = 0.05565 \pm 0.00031_{stat} \pm 0.00025_{syst}$\cite{Babusci:2014hxa} has a 0.72$\%$ accuracy, 
that is a factor $\simeq$ 5 better  with respect to the previous
measurement \cite{Chiang:1972rp}.

\vskip -0.6cm
\section{The charged kaon mass}
\label{Kmm}
The world average of the charged kaon mass, $M_{K^\pm}$, has a
precision of 13 keV~\cite{Agashe:2014kda} obtained mostly from two
precision measurements on kaonic atoms~\cite{Denisov:1991pu,Gall:1988ei} that differ by 60 keV or
4.6$\sigma$. A new precision measurement, at 10 keV level, is
desirable to reduce the bias that could arise from the average of the
previous results. The ongoing analysis on the KLOE sample of \Kcppp
aims to reach a precision of 5-7 keV using $\simeq$2.5 fb$^{-1}$ of
integrated luminosity. 
The \Kcppp channel is the best suited among kaon dominant decays for a precision measurement of
the mass due to the lowest Q--value (Q=75 MeV) that corresponds to the
best invariant mass resolution and minimal sensitivity to several
bias sources.
The candidates 
are selected through the same procedure applied for the
BR measurement (sec.\ref{BRm}), changing only the  requirements on the number of
secondaries at the decay vertex, from 2 to 3, to obtain the 3--pion
invariant mass, and on the vertex
position,  $\rho_{xy} \ge$~26~cm, 
to select decays inside the DC 
to avoid
biases from track extrapolation through the DC walls. 
On a MonteCarlo sample including all of the $\phi$ decays, to which we apply
the same procedure as on real data, we are studying 
some quality cuts, mostly to reduce the
invariant mass distribution tails: i) on the $\chi^2$ of the tracks
selected as pion candidates (currently $\chi^2 \le$~ 6); and ii) on the
difference between the pion momentum at the first hit in the DC and
its extrapolation at the decay vertex (currently
($|p_{\pi}^{firstHit}| \le |p_{\pi}^V|\le$~30~MeV).
On the same MC sample we have studied several functions and
fit procedures. We found good fit results using a function that
is the convolution of a Breit-Wigner with a gaussian distribution,
$\frac{dN}{M_{\pi\pi\pi}}=\frac{\gamma}{2\pi\sqrt{2\pi}\sigma}\int \!
\frac{e^{-\frac{(t-M_{\pi\pi\pi})^2}{2\sigma^2}}}{(t-M_{K^{\pm}})^2+\frac{\gamma^2}{4}}
\, \mathrm{d}t.$
Consistent results are obtained from the binned $\chi^2$ minimization and un-binned maximum--likehood
fit procedures.
\begin{figure}[!thb]
\centering
\vskip -0.4cm
\includegraphics[width=0.5\textwidth]{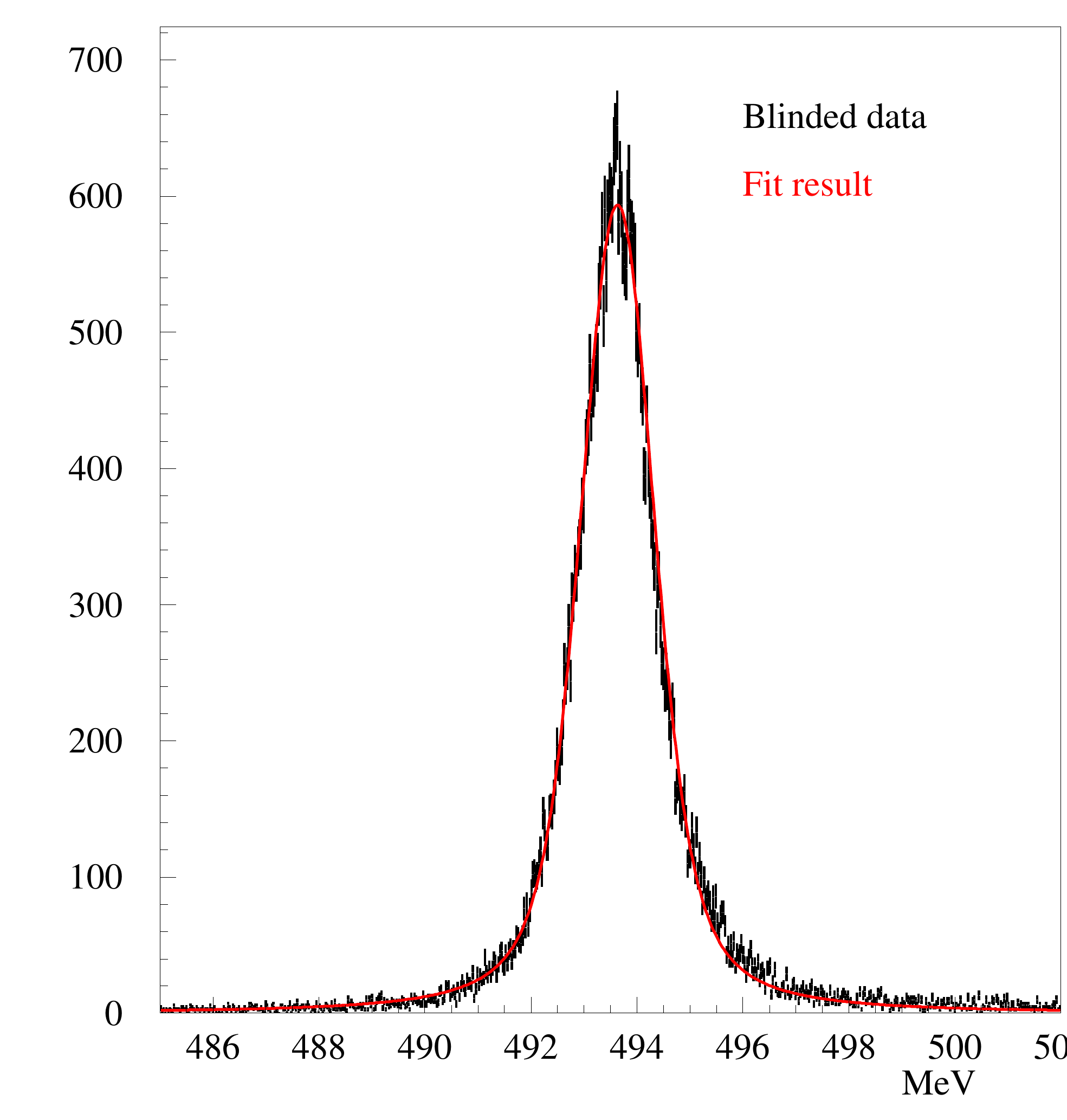} 
\vskip -0.3 cm
\caption[*]{The 3--pion invariant mass (M$_{\pi\pi\pi} + X$) (blinded
  data). The red curve is the result of the fit to the
convolution of  a Breit--Wigner with a gaussian
function.\label{fitdati}}
\vskip -0.2cm
\end{figure}

\noindent Besides the MonteCarlo sample of ($\simeq$23,000) \Kcppp decays
analyzed so far, a fast simulation program (toy--MC) has been developed
for the evaluation of effects such as  i) the presence of radiative decays; ii) bias and asymmetric
distribution of the reconstructed pion momentum, p$_\pi$, due for instance to the
treatment of energy loss and multiple scattering in the DC; iii)  an arbitrary
scale affecting p$_\pi$.
From these studies we can already rule out any bias on the charged kaon mass
from radiative decays. 
For the evaluation of the momentum scale factor $\alpha_p$, the analysis
of the charged kaon 2-body decays is in progress.  The scale factor
gives a bias to the kaon mass that depends on the Q--value of the decay channel,
bigger for K$_{\mu\nu}$ than for K$_{\pi\pi}$ and
K$_{\pi\pi\pi}$ decays. After correcting for other possible effects (point
ii) we can obtain $\alpha_p$ from the comparison  of the 2-body
decays. 
\noindent
 A blind analysis of the invariant mass distribution is in progress on
real data (fig.\ref{fitdati}).
We have obtained $\simeq$45,000 events after all the
analysis chain, with a negligible level of contamination that has been
evaluated by the MonteCarlo simulation, B/S $\simeq$~10$^{-4}$.

\section{Conclusions}
\par
We presented two recent KLOE results:  the CPT and Lorentz symmetry test, which constitutes the best limit in the quark sector of SME,
and the precision measurement of the 
branching fraction of the decay \Kp3pig .
Among the ongoing analyses, we discussed the status of the measurement for the precise determination of the charged kaon mass.
\par
As a future perspective,
the KLOE-2 experiment 
aims to continue and extend the 
physics program of its predecessor by collecting $\mathcal{O}(10\hbox{~fb}^{-1})$ of data at the upgraded DA$\Phi$NE with
an improved KLOE detector.
The KLOE-2 physics program has been described in detail in Ref.~\cite{prospects_kloe}
and among the main issues
includes $\ks$ physics, 
neutral kaon interferometry, 
tests of discrete symmetries and quantum mechanics,
test of the CKM unitarity, and other topics of kaon physics.
\par
Improvements are expected thanks to the increased luminosity and the better quality of reconstructed data.
The upgrade of the KLOE detector consists of the addition of (i) an inner tracker~\cite{kloeIT,morello}
 based on cylindrical GEM technology for the improvement of tracking and decay vertex resolution close to the interaction point (IP), (ii)
a $e^{\pm}$ tagging system~\cite{let,het,gauzzi} for the $\gamma\gamma$ physics, and (iii) two calorimeters~\cite{ref1,ref2,martini} in the final focusing region
to improve acceptance and efficiency for photons coming from the IP and neutral kaon decays
inside the detector volume.

\section{Acknowledgements}
\small\small We warmly thank our former KLOE colleagues for the access to the data collected during the KLOE data taking campaign.
We thank the DA$\Phi$NE team for their efforts in maintaining low background running conditions and their collaboration during all data taking. We want to thank our technical staff: 
G.F.~Fortugno and F.~Sborzacchi for their dedication in ensuring efficient operation of the KLOE computing facilities; 
M.~Anelli for his continuous attention to the gas system and detector safety; 
A.~Balla, M.~Gatta, G.~Corradi and G.~Papalino for electronics maintenance; 
M.~Santoni, G.~Paoluzzi and R.~Rosellini for general detector support; 
C.~Piscitelli for his help during major maintenance periods. 
This work was supported in part by the EU Integrated Infrastructure Initiative Hadron Physics Project under contract number RII3-CT- 2004-506078; by the European Commission under the 7$^{\mathrm{th}}$ Framework Programme through the `Research Infrastructures' action of the `Capacities' Programme, Call: FP7-INFRASTRUCTURES-2008-1, Grant Agreement No. 227431; by the Polish National Science Centre through the Grants No. 
2011/03/N/ST2/02641, 
2011/01/D/ST2/00748,
2011/03/N/ST2/02652,
2013/08/M/ST2/00323,
and by the Foundation for Polish Science through the MPD programme and the project HOMING PLUS BIS/2011-4/3.


%

\end{document}

%% file: definition.tex
\def\raE     {\rightarrow}

\def \Kp3pi  {${\rm K^+} \rightarrow \pi^+\pi^-\pi^+$~}
\def \Kcppp  {${\rm K^\pm} \rightarrow \pi^\pm\pi^-\pi^+$~}
\def \Kp3pig {${\rm K^+} \rightarrow \pi^+\pi^-\pi^+(\gamma)$~}

\newcommand{\braket}[2]{\ifm{\langle #1|#2 \rangle}}

\newcommand{\ks}          {\ensuremath{K_{S}}}
\newcommand{\kl}          {\ensuremath{K_{L}}}
\newcommand{\kp}          {\ensuremath{K^{+}}}
\newcommand{\kpm}          {\ensuremath{K^{\pm}}}
\newcommand{\kk}          {\ensuremath{K}}

\newcommand{\kzero}       {\ensuremath{K^{0}}}
\newcommand{\kzerob}      {\ensuremath{\bar{K}^{0}}}

\def\figb#1;#2;{\parbox{#2cm}{\epsfig{file=#1,width=#2cm}}}

\newcommand{\eV}{{e\kern-.07em V}}

\def\ifm#1{\relax\ifmmode#1\else$#1$\fi}
\def\pt#1,#2,{\ifm{#1\times10^{#2}}}

\def\Re{{\rm Re}}

\makeatletter
\newdimen\z@ \z@=0pt 
\newskip\z@skip \z@skip=0pt plus0pt minus0pt
\def\m@th{\mathsurround=\z@}
\def\ialign{\everycr{}\tabskip\z@skip\halign} 
\def\eqalign#1{\null\,\vcenter{\openup\jot\m@th
  \ialign{\strut\hfil$\displaystyle{##}$&$\displaystyle{{}##}$\hfil
    \crcr#1\crcr}}\,}
\makeatother

\def\sta#1;{\ifm{|\,#1\rangle}}    
\def\to{\ifm{\rightarrow}}

\def\bra#1;{\ifm{\langle\,#1\,|}}
\def\ket#1;{\ifm{|\,#1\,\rangle}}

\newcommand{\nc}{\newcommand}
\nc{\ChPT}{$\chi$PT}
\nc{\mrad}{\mbox{mrad}}
\nc{\mbo}{\mathversion{bold}}

\def\x{\ifm{\times}}   

\def\tt{\ifm{\widetilde{\tau}}}